\documentclass[aps,groupedaddress,10pt,amsmath,amssymb,twocolumn]{revtex4}
\usepackage[dvips]{graphics}
\usepackage{epsfig}
\usepackage{xspace}
\usepackage{amssymb,amsmath}

\begin{document}

\title{On the accuracy of DFT exchange-correlation functionals
for H bonds in small water clusters II: The water hexamer and van der Waals
interactions}
\author{Biswajit Santra$^1$}
\author{Angelos Michaelides$^{1,2}$}
\email{angelos.michaelides@ucl.ac.uk}
\author{Martin Fuchs$^1$}
\author{Alexandre Tkatchenko$^1$}
\author{Claudia Filippi$^{3}$}
\author{Matthias Scheffler$^1$}
\affiliation{$^1$Fritz-Haber-Institut der Max-Planck-Gesellschaft, Faradayweg 4-6, 14195 Berlin, Germany \\
$^2$Materials Simulation Laboratory, London Centre for Nanotechnology
and Department of Chemistry, University College London, London WC1E
6BT, UK\\
$^3$Instituut-Lorentz, Universiteit Leiden, P.O. Box 9506, NL-2300 RA
Leiden, and Faculty of Science and Technology and MESA+ Research
Institute, University of Twente, P.O. Box 217, 7500 AE Enschede,
The Netherlands}
\begin{abstract}
Second order M\o ller-Plesset perturbation theory (MP2) at the complete basis set (CBS) limit
and diffusion quantum Monte Carlo (DMC) are used to examine several low energy isomers of the water hexamer.
Both approaches predict the so-called ``prism'' to be the lowest energy isomer, followed
by ``cage'', ``book'', and ``cyclic'' isomers.
The energies of the four isomers are very similar, all being within
10-15 meV/H$_2$O.
This reference data is then used to evaluate the performance of
several density-functional theory (DFT) exchange-correlation (xc) functionals.
A subset of the xc functionals tested for
smaller water clusters [I: Santra \textit{et al.}, J. Chem. Phys. \textbf{127}, 184104 (2007)]
has been considered.
Whilst certain functionals do a reasonable job at predicting the absolute
dissociation energies of the various isomers (coming within 10-20 meV/H$_2$O),
\emph{none} predict the correct energetic ordering of the four isomers,
nor does any predict the correct low total energy isomer.
All xc functionals tested either predict the book or cyclic isomers
to have the largest dissociation energies.
A many-body decomposition of the total interaction energies within the hexamers
leads to the conclusion that the failure
lies in the poor description of van der Waals (dispersion) forces
in the xc functionals considered.
It is shown that the addition of an empirical pairwise
(attractive)
$C_{6}R^{-6}$ correction to certain functionals allows for
an improved energetic ordering of the hexamers.
The relevance of these results to density-functional simulations of liquid water is also briefly discussed.
\end{abstract}

\maketitle

\textbf{I. INTRODUCTION}\\

How good is density-functional theory (DFT) for hydrogen (H) bonds?
What is the best exchange-correlation (xc) functional for
treating H bonds?
Questions like these are far from uncommon for
developers and practitioners of Kohn-Sham DFT, particularly those interested in simulating collections
of atoms held together with H bonds.
Clearly imprecise and vague questions it is nonetheless important to answer them,
once, of course, terms like ``good'' and ``best'' have been defined
and consideration made to the properties of interest (energetic,
structural, dynamical, electronic).
Indeed considerable effort has been expended in an attempt to
answer questions like these
\cite{sms,Tsuzuki,Novoa,BenchmarkNonbonded,truhlar_pbe1w,Truhlar_hexamer,X3LYP_water,joel_pbe},
and with new xc functionals regularly appearing, there appears to be no end in sight
for such studies.\\

One particularly important class of H bonded systems, arguably the most
important, are the H bonds that hold water molecules together,
either as gas phase molecular clusters or condensed phase solid
(ice) and liquid water.
Kohn-Sham
DFT has been widely used to examine water under various
conditions and environments
\cite{sms,truhlar_pbe1w,Truhlar_hexamer,X3LYP_water,Todorova,
artacho,tuckerman,parrinello_water,nilsson_science,saykally_science_2004,
grossman_water,hu_michaelides_07,michaelides_appl_phys,Feibelman_science_2002,
ranea_michaelides,hu_michaelides_01,Enge_PRB_04}.
Along with this widespread application there have also been various benchmark
studies specifically aimed at accessing the performance of various xc functionals in
treating gas phase water clusters \cite{sms,truhlar_pbe1w,Truhlar_hexamer,X3LYP_water},
adsorbed clusters
\cite{hu_michaelides_07, michaelides_appl_phys, Feibelman_science_2002,
ranea_michaelides, hu_michaelides_01, Enge_PRB_04, michaelides_morgenstern,
AM_Farday_disc_07},
and liquid water
\cite{Todorova, artacho, tuckerman, parrinello_water, nilsson_science,
saykally_science_2004, grossman_water}.
In particular, the question of the performance of DFT xc functionals in describing
the structure and dynamics of liquid water has become a particularly
hot and contentious issue due to apparent discrepancies between
experiment and DFT
\cite{Todorova,artacho,tuckerman,parrinello_water,nilsson_science,
saykally_science_2004,grossman_water}.
Reconciling these differences, which are mainly concerned with
the radial distribution functions (RDFs) and diffusion
coefficient of liquid water,
remains an immensely important open question and is one
that is actively being addressed by many.
However, simultaneously addressing all the possible factors
which could account for the difference between the experimental
and theoretical RDFs and diffusion coefficients (e.g. quantum nuclear effects,
xc functional, density, basis
set, and so on) is far from straightforward and not particularly practicable.
Instead the course we and others have chosen to follow to shed light on
the performance of DFT xc functionals for treating water is
to investigate well-defined gas phase water clusters for which precise
comparison can be made to high level quantum chemistry calculations.
This approach allows the precise
performance limitations for a range of xc functionals
to be obtained, information that is likely to be of relevance to liquid water. \\

Previously we tested the performance of 16 xc functionals for the
equilibrium structures of the water dimer to pentamer
making reference to complete-basis set (CBS) extrapolated
MP2 data \cite{sms}.
That study revealed
that of the functionals tested the hybrid X3LYP \cite{X3LYP} and
PBE0 \cite{PBE0} functionals were the most accurate, both
coming within 10 meV/H bond of MP2 for each cluster.
Among the non-hybrid functionals mPWLYP \cite{mPW,LYP}
and PBE1W \cite{truhlar_pbe1w} offered the best performance \cite{sms}.
Here, we extend this work to the water hexamer.
The water hexamer is interesting and warrants
particular attention, not
least because it provides a critical test for DFT
xc functionals since there are four distinct isomers which lie
within 10-20 meV/H$_2$O of each other.
The isomers are known most commonly as the ``prism'', ``cage'',
``book'', and ``cyclic'' isomers (Fig. 1). Which one is
the lowest energy on the Born-Oppenheimer
potential energy surface with or without corrections for
zero point vibrations or the experimental ground state structure
at finite temperature has been a matter of debate for some time
\cite{tsai_jordan, kim_jordan_zwier_jacs, liu_saykally_nature,
gregory_clary_96, gregory_clary_97, kim_kim_hexamer, losada_leutwyler,
miller_science, Xantheas-2, hexamer_ccsdt, Truhlar_hexamer}.
For this paper we focus exclusively on the question of the lowest
total energy isomer without
zero point corrections, for which a consensus from
wave function based methods appears to have emerged recently in favor of the prism isomer
as being the lowest energy structure
\cite{kim_kim_hexamer, Xantheas-2, hexamer_ccsdt, Truhlar_hexamer}.
How many of the widely used xc functionals such as PBE, BLYP, and B3LYP perform for
the relative energies of these isomers remains unclear, although
there are indications that these and other DFT xc functionals are likely
to encounter problems for the hexamer
\cite{laasonen_parrinello, fitzgerald_hexamer, estrin, X3LYP_water}.
Other often cited reasons for being interested particularly
in water hexamers are that they represent a transition
from cyclic structures favored by smaller water clusters
to 3D structures favored by larger water clusters. And, that
water hexamers are believed to be important constituents
of liquid water and known to be building blocks
of various phases of ice \cite{ice_book}. \\

\begin{figure}
   \begin{center}
       \includegraphics[width=8cm]{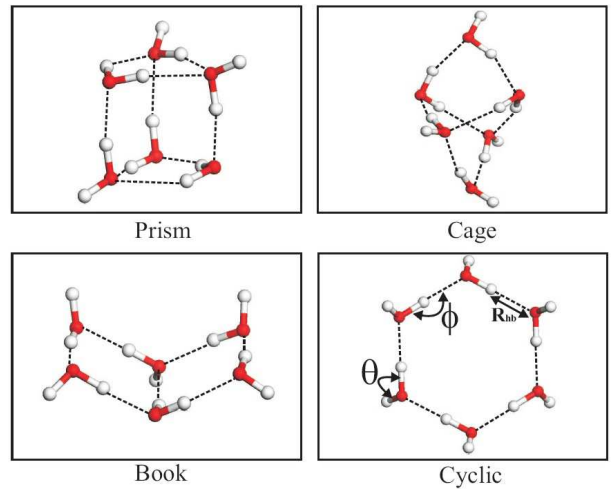}
   \caption{\label{fig1} Structures of the four isomers of the water
   hexamer considered here (obtained with MP2 and an aug-cc-pVTZ basis).
   The dashed lines indicate H bonds, with the conventional number of H bonds each cluster
   is assumed to have (prism = 9; cage = 8; book =7; and cyclic = 6) \cite{H_bond_comment}.
   Some of the structural parameters discussed in the text are included alongside
   the cyclic structure.}
   \end{center}
\end{figure}

In the following, we report a study in which the ability of several
popular xc functionals to describe the energies and structures of the four
water hexamers mentioned above is addressed.
Comparisons are made with reference data generated by ourselves with
2$^{nd}$ order M\o ller-Plesset perturbation theory (MP2)
at the complete basis set (CBS) limit
and diffusion quantum Monte Carlo (DMC).
The total energy ordering (i.e., neglecting zero point energies and finite temperature effects)
predicted by MP2 and DMC is the same and in the order
prism$<$cage$<$book$<$cyclic.
However,
all popular and widely used xc functionals tested fail
to predict the correct ordering of
the isomers, instead, they opt for
either the book or cyclic isomers as the lowest energy ones.
This discrepancy is largely attributed to the inability of DFT to correctly
capture the van der Waals (vdW) interaction between widely separated molecules in the clusters.
By including a semi-empirical $C_6R^{-6}$ correction
we are able to explain the origin of the failure of the tested xc functionals
and recover the correct energetic ordering
between the different conformers.\\

\textbf{II. METHODS AND REFERENCE DATA}\\

This paper involves the application of a variety of
theoretical approaches, which we now briefly describe.
Specifically, we discuss how the MP2 and DMC reference data is acquired,
and then the set-up for the DFT calculations.\\

\textbf{A. MP2}\\

MP2 has been used to compute structures and binding energies for each of the
four isomers.
All MP2 calculations have been performed with the Gaussian03 \cite{g03} and
NWChem \cite{nwchem} codes and all geometries were optimized with an
aug-cc-pVTZ basis set within the ``frozen core'' approximation i.e.,
correlations of the oxygen 1$s$ orbital were not considered
\cite{note_g03_nwchem}.
Although the aug-cc-pVTZ basis set is moderately large (92 basis
functions/H$_2$O), this finite basis set will introduce errors in
the predicted MP2 structures.
However, a test with the H$_2$O dimer reveals that the aug-cc-pVTZ
and aug-cc-pVQZ MP2 structures differ by only 0.004~\AA\ in the O-O
bond length and 0.16$^\circ$ in the H bond angle ($\phi$, Fig. \ref{fig1}).
Likewise, Nielsen and co-workers have shown that the MP2 O-O
distances in the cyclic trimer differ by 0.006~\AA\ between the
aug-cc-pVTZ and aug-cc-pVQZ basis sets with all other bonds
differing by $<$0.003~\AA\ \cite{Nielsen}.
For our present purposes these basis set incompleteness errors on
the structures are acceptable and it seems reasonable to assume that
the MP2 aug-cc-pVTZ structures reported here come with error bars
compared to the MP2/CBS limit of
$\pm$0.01~\AA\ for bond lengths and $\pm$0.5$^\circ$ for bond angles.
\\

Total energies and dissociation energies are known to be more
sensitive to basis set incompleteness effects than the geometries
are. To obtain reliable MP2 total energies and dissociation energies
we employ the aug-cc-pVTZ, aug-cc-pVQZ (172 basis functions/H$_2$O)
and aug-cc-pV5Z (287 basis functions/H$_2$O) basis sets in
conjunction with the well-established methods for extrapolating to
the CBS limit. Usually the extrapolation schemes rely on
extrapolating separately the Hartree-Fock (HF) and correlation
contributions to the MP2 total energy. For extrapolation of the HF
part we use Feller's exponential fit \cite{Feller}:
\begin{equation}
    E_X^{HF}=E_{CBS}^{HF}+Ae^{-BX} \quad ,
\label{eqn_HF_extrap}
\end{equation}
where, $X$ is the cardinal number corresponding to the basis set
($X$=3, 4, and 5 for the aug-cc-pVTZ, aug-cc-pVQZ, and aug-cc-pV5Z
basis sets, respectively).
$E_X^{HF}$ is the corresponding HF energy,
$E_{CBS}^{HF}$ is the extrapolated HF energy at the CBS limit, and
$A$ and $B$ are fitting parameters.
For the correlation part of the MP2 total energy we follow an
inverse power of highest angular momentum equation
\cite{Schwartz,Kutzelnigg,Corr-Fit}:
\begin{equation}
E_X^{Corr}=E_{CBS}^{Corr}+CX^{-3}+DX^{-5}\quad ,
\label{eqn_Correlation_extrap}
\end{equation}
where $E_{X}^{Corr}$ is the correlation energy corresponding to $X$,
$E_{CBS}^{Corr}$ is the extrapolated CBS correlation energy, and $C$
and $D$ are fitting parameters \cite{extrap_scheme_comment}.\\

\textbf{B. QUANTUM MONTE CARLO}\\

In order to assess the importance of correlation effects beyond the MP2 level,
we evaluated the binding energies of the water clusters using quantum Monte Carlo calculations (QMC).
QMC is a stochastic approach to solve the many-electron Schr\"odinger equation~\cite{foulkes01}.
The central quantity which determines the accuracy of a QMC calculation is the trial wave function,
i.e. a correlated ansatz for the many-electron wave function.
In variational Monte Carlo the expectation value of the many-electron Hamiltonian is computed as a statistical average
over a large number of electronic configurations which are sampled from the square of the trial
wave function using the Metropolis algorithm. An optimized trial wave function may be obtained
within variational Monte Carlo based on variational principles for the variance of the local energy or the energy~\cite{umrigar05}. This trial wave function is then used in diffusion Monte
Carlo (DMC) which yields the best energy within the fixed-node approximation (i.e.
projecting out the lowest-energy state with the same nodes as the trial wave function).\\

DMC calculations yield highly accurate results for a wide variety of chemical systems
(molecules and solids) and properties (binding energies, reaction energetics) as
shown, for example, in Refs.~\cite{foulkes01,manten01,grossman02,healy02}.
Recent studies of hydrogen-bonded (and stacked aromatic) molecular dimers~\cite{needs07,korth08,fuchs05,sorella07,zaccheddu08}
demonstrate that DMC describes the interaction energies of such non-covalently bonded systems
in very close agreement with the best available CCSD(T)/CBS estimates, in particular where corrections beyond
MP2/CBS are significant~\cite{s22}.\\

In the present work, the trial wave functions are chosen of the Slater-Jastrow
form with $\Psi = D_{\uparrow}D_{\downarrow}e^J$, i.e. as the product of Slater determinants $D_{\sigma}$
of one-particle orbitals for the spin-up and spin-down electrons and a Jastrow correlation factor $e^J$
depending on the electron-electron and electron-nucleus distances~\cite{filippi96}.
The one-electron orbitals are represented in an atomic Gaussian basis and generated from DFT-B3LYP calculations
using the GAMESS code~\cite{gamess}. The parameters of the Jastrow correlation factor are
optimized using the variance minimization method~\cite{umrigar88}.
The atomic cores are represented using the nonlocal pseudopotentials of Ref.~\cite{burkatzki07}
and included in diffusion QMC within the usual localization approximation
(i.e. nonlocal potentials are transformed into local operators by projection onto the trial wave function).
The QMC calculations are performed using the CHAMP package~\cite{champ,computation}. All QMC results
reported below are from DMC. The results for the different isomers are given in Table \ref{tb1}
and calculated for MP2/aug-cc-VTZ geometries (see Sec.~II.A) using a time step of
$\tau = \frac{1}{80}$ $E_h^{-1}$ and a target population of 800~walkers.\\

The main approximations in our DMC calculations are the fixed-node and pseudopotential
localization approximations, and the quality of both is determined by the choice of the
trial wave function. To estimate their effect on the hydrogen bond energies we have
carefully analyzed the results of DMC calculations for the water dimer (in terms of
the choice of pseudopotentials, basis set, terms in the Jastrow factor, and the time
step in DMC) using the same form of the trial wave function as for the water hexamers.
For the dissociation energy of the water dimer we obtain $D_e=218\pm 3$~meV, in agreement
with the recent result of Gurturbay and Needs (GN), $218\pm 3$~meV, obtained
by a DMC calculation that employed a different set of pseudopotentials but appears
otherwise essentially analogous to ours~\cite{needs07}. These pseudopotential DMC results
are consistent with the CCSD(T)/CBS result of Ref.~\cite{klopper00} ($217.7$~meV)
and the all-electron DMC result of GN ($224\pm 4$~meV, for a Slater-Jastrow wave function using DFT-B3LYP orbitals).
GN furthermore showed that going beyond the localization approximation for
nonlocal pseudopotentials produces equivalent results for the water dimer dissociation energy to within $5$~meV.
As the nodes of the trial wave function are given by its determinantal part, i.e. by $D_\sigma$,
we also use orbitals from HF instead of DFT-B3LYP
calculations to build the Slater determinants and thus provide a test of the
sensitivity of the DMC results to changes in the nodes. While HF
orbitals noticeably increase the total energies of the monomer and dimer compared to DFT-B3LYP orbitals,
we find that these changes cancel in the DMC dissociation energy. Our DMC-HF
result is $214\pm 6$~meV compared to $218\pm 8$~meV for all-electron DMC-HF~\cite{needs06}.
On the other hand, GN showed that the inclusion of so-called backflow correlations
to alter the nodes in a Slater-Jastrow wave function produces a slightly
stronger hydrogen bond, changes being $< 20$~meV in their pseudopotential DMC
and somewhat smaller in their all-electron DMC calculation.
From the above comparison of our DMC results for the water dimer with the best
available theoretical reference data, DMC and CCSD(T), errors in the
hydrogen bond strength due to the fixed-node and pseudopotential
(localization) approximation appear small. We therefore expect that
our DMC calculations provide an accurate account of the interactions
between water molecules also in the hexamers, i.e. within $\approx 10$~meV/H$_2$O.
To further corroborate this estimate requires additional investigation, in particular
of the accuracy of the available, different pseudopotentials as well as of refinement
of the trial wave functions. This is beyond the scope of the present study, but we note that previous
DMC-HF studies using different pseudopotentials (and slightly different geometries)
than employed here found somewhat larger dissociation energies of the water dimer
($245\pm 9$~meV ~\cite{luechow05} and $232\pm 4$~meV~\cite{korth08}) than in the
present work and in Ref.~\cite{needs07}. The
pseudopotentials used in these and the present study are both based on
atomic HF calculations, yet their functional form is different.
The accuracy of the pseudopotentials used here has
been explicitly demonstrated for molecular properties of diatomic molecules
at the CCSD(T) level~\cite{burkatzki07}.\\

\textbf{C. KOHN-SHAM DFT}\\

We have performed DFT calculations with 12 different xc functionals,
chosen either because they are popular or have previously
been shown to perform well for the strengths
of hydrogen bonds between water molecules.
Specifically, we have examined the following generalized gradient
approximation (GGA) functionals: PW91 \cite{PW91}, PBE \cite{PBE},
PBE1W \cite{truhlar_pbe1w}, mPWLYP \cite{mPW,LYP}, BP86 \cite{Becke88,Perdew86},
BLYP \cite{Becke88,LYP}, and XLYP \cite{X3LYP}).
The meta-GGA TPSS \cite{TPSS} has also been considered as well as the following
hybrid functionals: PBE0 \cite{PBE0}, X3LYP \cite{X3LYP},
B3LYP \cite{B3LYP-1,B3LYP-2,B3LYP-3,LYP}, and B98 \cite{B98}.
The local-density approximation (LDA) has not been considered
since it is known to overestimate the dissociation energy of
water clusters by $>$50\% \cite{fitzgerald_hexamer}.\\

Most DFT calculations have been performed with the Gaussian03 \cite{g03} and
NWChem \cite{nwchem} codes.
Such calculations are all-electron and employ Gaussian-type orbital basis sets.
Geometries were optimized with an
aug-cc-pVTZ basis set and energies with an aug-cc-pV5Z basis set.
We have shown before that such large basis sets are, for DFT,
sufficiently large to reflect the true performance of each xc
functional at a level of accuracy that is reasonably expected to
approach the basis set limit to within about 0.5 meV/H bond or
better \cite{sms}.\\
%

While standard quantum chemistry software packages, such as the ones mentioned
above, can be conveniently used for the simulation of small water clusters, one of
our longer term goals is the accurate simulation of condensed phases of water such
as ice or liquid water. Therefore, we have also performed selected DFT
calculations with codes suitable for condensed phase simulations, such as
the plane-wave pseudopotential code CPMD \cite{cpmd} and the all electron
numeric atom-centered orbitals (NAO) code FHI-aims, which originates from our laboratory \cite{FHI-aims}.
A byproduct of such effort is the interesting comparison of three different methodologies
for DFT calculations (Gaussians, plane-waves, and NAOs) of the energetics
of hydrogen bonded systems.
For the pseudopotential plane-wave DFT calculations in CPMD we have used hard
Goedecker \emph{et al.} \cite{godecker_pp,krack_pp}
pseudopotentials along with an energy cutoff of at least 200 Ry
for the plane-wave kinetic energy.
For each hexamer an appropriate cell size was chosen to leave
at least 10 {\AA} of vacuum on each side of the cluster.
This cell size was found to be converged by
performing selected simulations with a larger vacuum of 15 {\AA}
and with or without the Hockney Poisson solver \cite{hockney}
for electrostatic decoupling between neighboring cells.
In the case of the
all-electron
NAO calculations with the FHI-aims code
we have employed
hydrogenic basis functions and carefully benchmarked all calculation parameters
(basis set, grid size, cutoff potential) to achieve extreme convergence equivalent
to or better than an aug-cc-pV5Z Gaussian basis set.
As we will discuss below, a comparison between these three methods on the energetics of
small water clusters (dimer-pentamer)
reveals that the differences in the binding energy are on the order of 0.1 meV between
Gaussian and FHI-aims and of 1 meV between Gaussian and CPMD,
a value which is negligible for all our conclusions.
Further, we have implemented a $C_6R^{-6}$ semi-empirical correction
for vdW interactions both in CPMD
and FHI-aims,
so that we could perform full geometry optimizations with and without this correction.\\

\textbf{D. DISSOCIATION ENERGY}\\

For MP2, DFT, and DMC we have calculated dissociation
energies per H$_2$O $(D_{e}^n)$ which is given by,
\begin{equation}
    D_{e}^n = (E^{nH_2O} - nE^{H_2O})/n_{H_2O} \quad ,
\label{eqn_disso_energy}
\end{equation}
where $E^{nH_2O}$ is the total energy of each cluster with $n$
H$_2$O molecules, $E^{H_2O}$ is the total energy of a H$_2$O
monomer, and $n_{H_2O}$ is the number of water molecules in the cluster.
\\
\begin{table*}
\caption {\label{tb1}
Dissociation energies
of the four water hexamers obtained
from various electronic structure approaches:
MP2/CBS;
diffusion quantum Monte Carlo (DMC);
CCSD(T) with a triple-$\zeta$ basis set (Ref. \cite{hexamer_ccsdt}); 12 different
DFT exchange-correlation functionals computed, unless indicated otherwise, with an
aug-cc-pV5Z basis set; and HF at the CBS limit.
The most stable isomer
from each method is indicated in bold and
the relative energies of the other isomers are given in parenthesis.
Mean errors (ME) and mean absolute errors (MAE) in dissociation energies, averaged
over the four hexamers in comparison with MP2 and DMC are also given.
All structures were optimized consistently with MP2, HF
and each DFT functional with an aug-cc-pVTZ basis set
except for the DMC calculations which used the MP2 structures.
DFT xc functionals are arranged here with increasing value of MAE from MP2.
All values are in meV/H$_2$O (1kcal/mol = 43.3641 meV).}
\begin{ruledtabular}
\begin{tabular}{c|cccc|cc|cc}
&&&&& \multicolumn{2}{c|}{MP2} & \multicolumn{2}{c}{DMC}\\
\hline
Method &  Prism &     Cage &    Book &    Cyclic & MAE & ME & MAE  & ME\\
\hline
MP2 &\textbf{332.3}   & 331.9 (0.4)  &  330.2 (2.1)  & 324.1 (8.2)& --- & --- & --- & ---\\
DMC\footnote{the statistical errors on the dissociation energies of
prism, cage, book, and cyclic are
$\pm$1.0, $\pm$0.9, $\pm$1.0, and $\pm$1.0 meV/H$_2$O, respectively.
For the relative energies of the cage, book, and cyclic with
respect to the prism (calculated as differences of total energies
of the isomers rather than their dissociation energies) the
statistical errors are
$\pm 1.0$, $\pm 1.1$, and $\pm 1.1$~meV/H$_2$O, respectively.}
&\textbf{331.9} & 329.5 (2.4)& 327.8 (4.1)& 320.8 (11.1) & --- & --- & ---& ---\\
CCSD(T)\footnote{Reference \cite{hexamer_ccsdt}}
   & \textbf{347.6}   & 345.5 (2.1)  & 338.9 (8.7)& 332.5 (15.1) & --- & --- & --- & ---\\
PBE0     &322.9 (8.0)   & 325.3 (5.7) &  \textbf{330.9}  & 330.8 (0.1) & 5.9 & -2.1 & 6.6 & 0.0\\
mPWLYP   &323.2 (10.4)  & 325.9 (7.7) &  \textbf{333.6}  & 333.3 (0.3)& 6.9 & -0.6 & 7.7 & 1.5\\
X3LYP    &317.2 (8.8)  & 319.2 (6.8) &  325.8 (0.2) & \textbf{326.0} & 8.5 & -7.6 & 7.1 & -5.5\\
PBE1W    &315.2 (6.9) & 314.8 (7.3) &  \textbf{322.1}  & 321.5 (0.6)& 11.3 & -11.3 & 9.5 & -9.1\\
PBE      &336.1 (9.5) & 339.4 (6.2) &  \textbf{345.6}  & 344.1 (1.5) & 11.7 & 11.7 & 13.8 & 13.8\\
%
%
B98      &305.3 (7.3) & 306.8 (5.8) &  \textbf{312.6}  & 312.5 (0.1) & 20.4 & -20.4 & 18.2 &-18.2\\
TPSS     &303.9 (12.8)  & 302.8 (13.9) &  313.6 (3.1) & \textbf{316.7} & 20.4 & -20.4 & 18.3 &-18.3\\
PW91     &351.4 (10.2)  & 354.7 (6.9) &  \textbf{361.6}  & 360.3 (1.3) & 27.3 & 27.3 & 29.5& 29.5\\
BP86     &294.9 (13.6)  & 297.4 (11.1) &  \textbf{308.5}  & 306.6 (1.9) & 27.8 & -27.8 & 25.7& -25.7\\
B3LYP    &294.4 (12.3)  & 297.1 (9.6) &  305.1 (1.6) & \textbf{306.7} & 28.8 & -28.8 & 26.7 & -26.7\\
XLYP     &287.9 (10.0)  & 286.9 (11.0)  & 296.3 (1.6)  & \textbf{297.9} & 37.4 & -37.4 & 35.3 & -35.3\\
BLYP     &273.6 (16.2)  & 277.4 (12.4) & 287.5 (2.3)  & \textbf{289.8} & 47.6 & -47.6 & 45.4 & -45.4\\
BLYP\footnote{Full geometry optimization was done with FHI-aims code.}
& 273.6 (16.2)& 277.3 (12.5)& 287.4 (2.4)& \textbf{289.8}& 47.6& -47.6& 45.5& -45.5 \\
BLYP\footnote{Full geometry optimization was done with CPMD code.
}
& 272.1 (17.6)& 276.0 (13.7)& 286.7 (3.0)& \textbf{289.7}& 48.5 & -48.5 & 46.4 & -46.4 \\
HF  &222.9 (12.2) & 224.4 (10.7) & 230.6 (4.5)  & \textbf{235.1} & 101.4 & -101.4 & 99.3 & -99.3 \\
\end{tabular}
\end{ruledtabular}
\end{table*}

\textbf{III. RESULTS}\\

Now we present and discuss our MP2 and DMC reference data. Following this we evaluate the
accuracy of the 12 exchange-correlation functionals considered, and then present a many-body decomposition of
the total dissociation energies as well as a detailed discussion of the value of
accounting for vdW dispersion forces in these clusters.\\

\textbf{A. REFERENCE DISSOCIATION ENERGIES}\\

Following the procedure outlined above, we obtain
MP2 dissociation energies at the CBS limit for the prism, cage, book, and
cyclic hexamers of 332.3, 331.9, 330.2, and 324.1 meV/H$_2$O,
respectively (see Table \ref{tb1})
\cite{mp2_error}.
Thus with MP2 the prism is the most stable structure and the energetic
ordering of the isomers is prism$<$cage$<$book$<$cyclic.
We note that this is consistent with
the previous MP2/CBS study of the water hexamer reported by Xantheas
\emph{et al.} \cite{Xantheas-2,note_extrap}.
The DMC calculations also find the prism to be the most stable isomer and predict the same energetic ordering as MP2.
Clearly, the cyclic is the least stable isomer while the
prism and the cage isomers appear energetically very close as
they only differ by about two standard errors.
Moreover,
the absolute dissociation energies obtained with
DMC and MP2 are within 4 meV/H$_2$O of each other for all four clusters (Table I).
The sequence prism$<$cage$<$book$<$cyclic is also consistent with recent CCSD(T)
calculations \cite{hexamer_ccsdt,Truhlar_hexamer},
although the absolute binding energies from CCSD(T) when reported
\cite{hexamer_ccsdt} are some 10-15 meV/H$_2$O
larger than our MP2/CBS dissociation energies.
Most of this difference can, however, be attributed to the smaller
(aug-cc-pVTZ) basis set used in the CCSD(T) study \cite{note_CCSD(T)}.
Therefore, it is clear that all the explicitly correlated wave function based
methods [MP2, DMC, CCSD(T)] predict the
same low energy structure -- prism -- and the same
energetic ordering: prism$<$cage$<$book$<$cyclic.
With this consensus from different methods it now seems that the question
of which isomer is the lowest energy on the Born-Oppenheimer potential energy surface
(in the absence of contributions from zero point vibrations) is resolved in
favor of the prism, and that
suggestions to the contrary are not correct \cite{X3LYP_water}.
There remain, of course, minor differences in the relative
energetic ordering of some structures on the order of 5 meV/H$_2$O
[notably CCSD(T) predicts particularly unstable book and cyclic structures compared to MP2,
with DMC being in between].
Resolving such small remaining differences is beyond the scope of the current paper,
which instead now focuses on how the various DFT functionals do in describing the energies
and structures of these clusters.\\

\textbf{B. DFT DISSOCIATION ENERGIES}\\

We turn now to the results obtained with the various DFT xc functionals
and first consider: (i) if the DFT xc functionals tested are able to
predict the correct energetic ordering of the
four hexamer isomers; and (ii) what are the absolute errors in the total
dissociation energies for each of the isomers.
The answer to the first question is simple.
All popular and widely used functionals tested fail to predict
the correct minimum energy isomer.
Instead of identifying the prism as the minimum energy conformer,
all xc functionals tested either opt for the \emph{cyclic} or \emph{book}
conformers (Table \ref{tb1}).
This includes the X3LYP and PBE0 functionals, which,
in our previous study \cite{sms}, were identified as the most accurate xc functionals
of those tested on the global minimum structures of small water clusters.
%
%
It is somewhat discouraging that most of the xc functionals
tested despite being immensely popular for liquid water
simulations, fail to predict the correct low energy structure for a system as seemingly
simple as six water molecules.
However, the failure is not entirely unexpected given that according to
the wave function methods all four structures are so close in
energy (within 10-15 meV/H$_2$O).\\

With regard to the second issue of how well the functionals perform
at predicting the absolute binding energies of the clusters, the
best functionals are PBE0, mPWLYP, and X3LYP, producing mean absolute
errors (MAE) averaged over the four clusters of 6, 7, and 9 meV/H$_2$O.
PBE and PW91 produce errors of 12 and 28 meV/H$_2$O,
respectively.
%
B98 and TPSS both have a MAE of 20 meV/H$_2$O.
B3LYP and BLYP under-bind by $\sim$29 and $\sim$48 meV/H$_2$O,
respectively.
%
All of these conclusions are largely consistent with our previous
study on smaller water clusters \cite{sms}.
We note that the MAEs discussed are those obtained
with respect to the MP2/CBS reference data.
If instead we use the DMC results as the reference, the conclusions
all remain essentially the same. This can be seen from Table \ref{tb1},
and is, of course, due to the fact that the DMC and MP2 reference data is
so similar (always within 4 meV/H$_2$O).\\

\begin{figure*}
   \begin{center}
       \includegraphics[width=14cm]{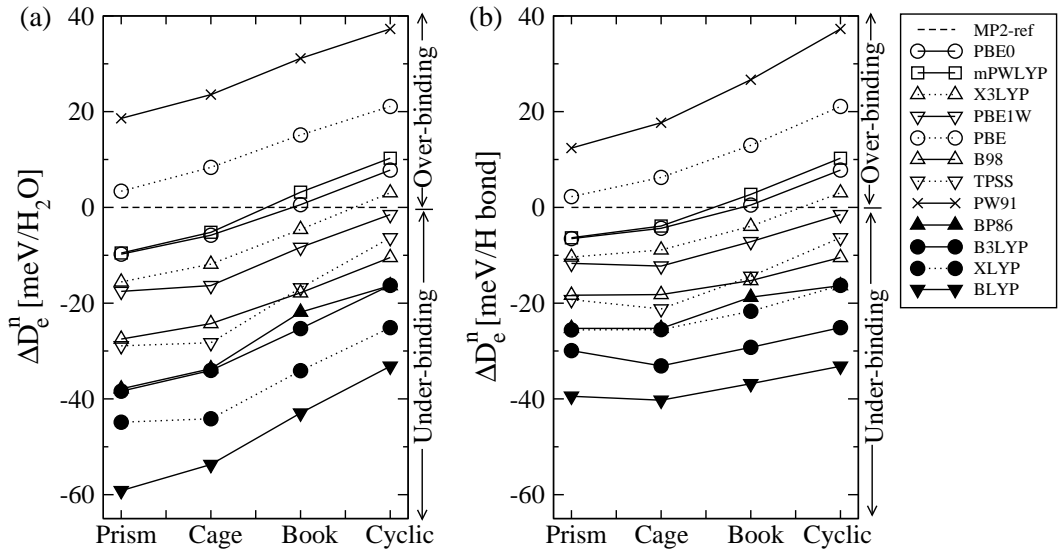}
   \end{center}
    \caption{\label{fig2} Difference in the dissociation energy
     $(\Delta\text{D}^\text{n}_\text{e})$ in (a) meV/H$_2$O and (b) meV/H bond
     between the various DFT xc functionals and MP2.
     In (b) the generally accepted number of H bonds in the prism, cage, book, and cyclic
     isomers of 9, 8, 7, and 6, respectively, have been used \cite{H_bond_comment}.
     Positive values correspond to an over-estimation of the dissociation
     energy by a given DFT xc functional.
     We note that the reference MP2 dissociation energies are at the CBS limit
     whereas for the DFT xc functionals an aug-cc-pV5Z basis set has been employed.
     Lines are drawn to guide the eye only.}
\end{figure*}

Looking more closely at how the functionals perform for specific
clusters, we have plotted in Fig. \ref{fig2}(a) and \ref{fig2}(b)
the difference
between each functional and MP2/CBS  ($\Delta D_e^n$) for all four isomers.
Since each cluster nominally has a different number of H bonds \cite{H_bond_comment}, and we are interested also
in the description of H bonds,
in Fig. \ref{fig2}(b) we also plot the error per H bond for each of the clusters.
%
%
%
Fig. \ref{fig2} proves to be very illuminating and from it we extract
the following key conclusions:
(i) Upon moving from the prism to the cyclic isomer (as
plotted in Fig. \ref{fig2}), all xc
functionals display a trend towards increased binding;
(ii) Most functionals underbind the prism, with PBE and PW91 being the only
exceptions;
(iii) As we saw before for the dimer to pentamer \cite{sms}, here also BLYP performs consistently
when we consider the error per H bond, coming around $\sim$35 meV/H bond
off MP2. Likewise XLYP yields very similar errors
for all four isomers when considered on a per H bond basis.
We will draw upon
these conclusions later.\\

Another interesting finding is that the calculations on different
water hexamers agree within 0.1 meV/H$_2$O between the all-electron Gaussian03 and
FHI-aims codes and within 1.5 meV/H$_2$O between Gaussian03 and the pseudopotential
plane-wave CPMD code (Table \ref{tb1}).
The latter value is most probably due to the difference in treatment
of core electrons, however this difference is still very small for all practical purposes.
This level of agreement is also achieved for the smaller clusters -- dimer to pentamer --
in their equilibrium geometries \cite{note_on_small_clusters}.
This again reinforces that the basis sets employed here are
sufficiently large to reflect the true performance of a given
xc functional, absent of basis set incompleteness errors.\\

\begin{table*}
\caption {\label{tb2} Mean absolute error (MAE) of the various DFT
functionals from MP2 for five different structural parameters,
averaged over the four water hexamers examined here. The numbers in
bold all have MAE $\leq$0.010 {\AA} for bond lengths and
$\leq$0.50$^\circ$ for bond angles. Mean errors (ME) are given in
parenthesis. MP2 and DFT (and HF) structures were optimized consistently with MP2 and
with each DFT functional (and HF) with an aug-cc-pVTZ basis set. The DFT+vdW structures were
optimized with a numerical atom-centered basis set (FHI-AIMS code). The order of
the DFT xc functionals is the same as in Table I.}
\begin{ruledtabular}
\begin{tabular}{clllll}
& $\Delta \text{R}_\text{{O-O}}$ (\AA)& $\Delta \text{R}_\text{{hb}}$ (\AA)& $\Delta \text{R}_\text{{O-H}}$ (\AA)
& $\Delta\phi$  $(^{\circ})$ & $\Delta\theta$  $(^{\circ})$ \\
\hline
PBE0  & 0.023 (-0.017) & 0.028 (-0.018) & \textbf{0.002} (0.000) & 0.96 (-0.01)& 0.69 (+0.69) \\
mPWLYP& 0.021 (+0.021) & 0.019 (+0.008) & 0.013 (+0.013)& 0.95 (-0.25)& \textbf{0.49} (+0.49) \\
X3LYP & \textbf{0.009} (+0.008) & 0.012 (+0.009) & \textbf{0.000} (0.000) & \textbf{0.48} (-0.29)& 0.98 (+0.98) \\
PBE1W & 0.062 (+0.045) & 0.096 (+0.051) & 0.011 (+0.011)& 3.98 (-0.64)& \textbf{0.32} (+0.03) \\
PBE   & 0.032 (-0.019) & 0.055 (-0.036) & 0.014 (+0.014)& 1.87 (+0.12)& \textbf{0.24} (+0.18) \\
PBE+vdW & 0.026 (-0.022) & 0.044 (-0.039) & 0.012 (+0.012) & 1.11 (+0.26) & 0.21 (+0.03)\\
B98   & 0.025 (+0.025) & 0.028 (+0.028) & \textbf{0.001} (-0.001)& 1.07 (-0.20)& 0.66 (+0.66) \\
TPSS  & 0.094 (+0.040) & 0.155 (+0.058) & 0.011 (+0.011) & 6.03 (-0.87)& 0.58 (0.53)           \\
PW91  & 0.039 (-0.034) & 0.060 (-0.051) & 0.014 (+0.014)& 1.59 (+0.15)& \textbf{0.36} (+0.33) \\
BP86  & 0.032 (-0.026) & 0.055 (-0.046) & 0.016 (+0.016)& 1.65 (+0.27)& \textbf{0.28} (+0.16) \\
B3LYP & 0.019 (+0.019) & 0.020 (+0.020) & \textbf{0.000} (+0.000) & 0.61 (-0.28)& 0.89 (+0.89) \\
XLYP  & 0.092 (+0.082) & 0.113 (+0.091) & 0.011 (+0.011)& 3.73 (-0.99)& 0.52 (+0.52) \\
BLYP  & 0.039 (+0.039) & 0.029 (+0.028) & 0.012 (+0.012)& 1.29 (-0.20)& \textbf{0.39} (+0.39) \\
BLYP+vdW & 0.030 (-0.026) & 0.052 (-0.044) & 0.013 (+0.013)& 1.94 (0.59)& \textbf{0.63} (+0.63)\\
%
HF & 0.163 (+0.163) & 0.199 (+0.199) & 0.026 (-0.026)& 1.64 (-0.929)& 1.62 (+1.62) \\
\end{tabular}
\end{ruledtabular}
\end{table*}

\begin{figure*}
   \begin{center}
       \includegraphics[width=12cm]{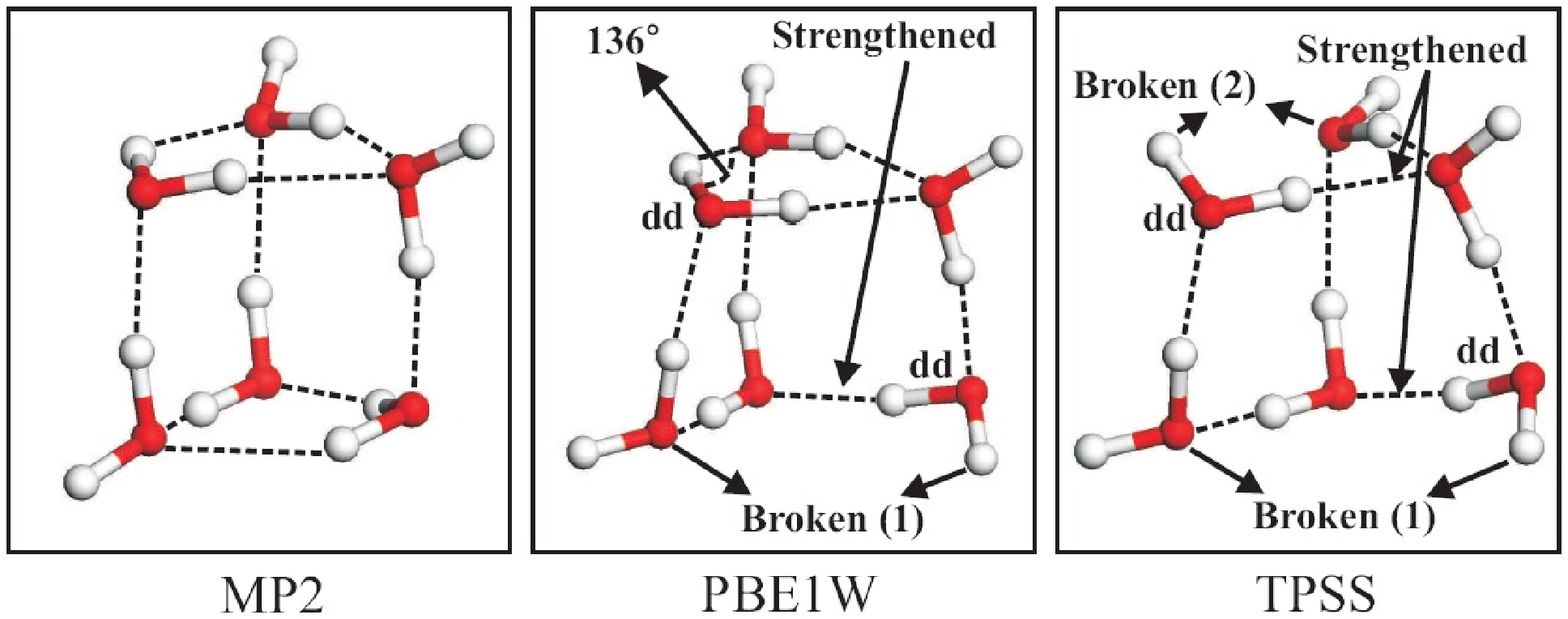}
   \end{center}
   \caption{\label{fig3} Structures of the prism isomer optimized with MP2
   and the PBE1W and TPSS xc functionals. Dashed lines indicate
   H bonds. For PBE1W one H bond is broken and for TPSS two H bonds
   are broken, each broken H bond being associated with a double donor (dd) water molecule.
   The other H bonds which get stronger as a result of the bond breaking are
   also indicated.
   A very bent H bond angle of 136$^\circ$ is also shown in the upper
   triangle of the PBE1W structure.}
\end{figure*}

\textbf{C. GEOMETRY}\\

Let us now consider the quality of the geometrical predictions made by the
various xc functionals.
The five key structural
parameters of the H$_2$O clusters (some of them are shown in
Fig. \ref{fig1}) that we evaluate are:
(i) The distance between adjacent oxygen atoms involved in a H bond,
$\text{R}_\text{{O-O}}$;
(ii) The length of a H bond, given by the distance between the donor
H and the acceptor O, $\text{R}_{\text{O}\cdots \text{H}}=
\text{R}_{\text{hb}}$ (Fig. \ref{fig1});
(iii) The H bond angle, $\angle{(\text{O}\cdots \text{H-O})}= \phi$
(Fig. \ref{fig1});
(iv) The internal O-H bond lengths of each water,
$\text{R}_{\text{O-H}}$; and
(v) The internal H-O-H angle of each water,
$\angle({\text{H-O-H}})=\theta$ (Fig. \ref{fig1}).\\

In Table \ref{tb2}, the MAE and ME
of each xc functional compared to MP2 and averaged
over all four clusters are reported.
This provides a broad overview of how each functional performs,
revealing that for structural predictions X3LYP
is the most accurate functional.
X3LYP outperforms all other functionals for almost all structural
parameters considered with an average error of only 0.02 \AA\ for the bond lengths
and 0.5$^\circ$ for the bond angles.
Considering the predicted O-O distances, on average, X3LYP, mPWLYP,
PBE1W, TPSS, B98, B3LYP, BLYP, and XLYP
predict slightly longer (0.008 to 0.082 \AA) distances, whereas, PBE0, PBE, BP86,
and PW91 produce slightly shorter O-O distances (0.017 to 0.034 \AA).
This conclusion also holds for the related quantity $\text{R}_{\text{hb}}$.
For the O-H bond length, $\text{R}_{\text{O-H}}$, on average all functionals
perform reasonably well coming within 0.02 \AA. In particular the results for X3LYP,
PBE0, B98, and B3LYP are nearly identical to MP2.
For the internal H-O-H angle $\theta$, the MAE from all the functionals
is within $\sim$1.0$^\circ$.
Finally, for the H bond angle, $\phi$, X3LYP, B3LYP, PBE0, and mPWLYP
perform the best, all coming within 1.0$^\circ$.
For this quantity, however, several functionals exhibit quite large discrepancies.
%
Specifically, XLYP, PBE1W, and TPSS yield
average MAEs of 3.7$^\circ$, 3.9$^\circ$, and 6.0$^\circ$, respectively.
As we go from cyclic to book to cage to prism, the H bond angles in the clusters
become increasingly non-linear (179$^\circ$ for cyclic, $\sim$160$^\circ-170^\circ$
for book, $\sim$152$^\circ-166^\circ$ for cage, and $\sim$135$^\circ-168^\circ$ for prism)
and it appears that certain xc functionals struggle to reliably
describe such non-linear H bonds.
Indeed closer inspection reveals that the largest errors in $\phi$ are encountered
for the prism isomer.
In this isomer there are two water molecules that are each involved in donating
two hydrogen bonds (the molecules labeled \emph{dd} for double donor in Fig. \ref{fig3}),
and according to MP2 the H bonds these molecules donate are very bent (i.e., values of $\phi$ $\sim$135$^\circ$).
Several of the xc functionals fail to describe these very non-linear essentially
putative H bonds, and
for one or both of the waters in the prism sacrifice a single very non-linear H bond to enable the other
to become more linear and hence stronger (Fig. \ref{fig3}).
TPSS fails for both double donor water molecules and PBE1W and XLYP fail to describe one of them.
The limitations of functionals such as those considered here in
describing non-linear putative H bonds in water clusters has
also recently been pointed out by Shields and Kirschner \cite{shields_kirschner}.
There it was argued that vdW dispersion forces are critical
to the binding of such weak H bond structures.
We tend to agree with this conclusion and will show more evidence in support
of it below.\\ \\

\textbf{D. MANY-BODY DECOMPOSITION OF THE DISSOCIATION ENERGIES}\\

To identify precisely where the problem with the
DFT xc functionals lies in correctly describing the energetic
ordering of the various isomers, we have performed a many-body
decomposition of the total dissociation energies of the hexamers.
This has involved decomposing the total interaction energy within
the clusters into 1-body, 2-body, $\cdots$, 6-body contributions.
Such many-body expansions have before proved useful in understanding
the binding in H bonded clusters (including water clusters). A full
description of the procedure involved can be found in Refs.
\cite{xantheas_many_body, xantheas_cooperativity, pedulla_vila_jordon, quack_hf}.
Very briefly, the total 1-body energy
is the energy
cost incurred upon deforming all six monomers from the equilibrium isolated monomer structure to
the structures they assume in a given hexamer.
The total 2-body interaction energy is the sum of all possible dimer
interactions within the hexamer, i.e., the total energy (gain) to form all
possible water dimers within a given hexamer from each of its (deformed) monomers.
The total 3-body interaction corresponds to the energy (gain) to form all possible
trimer combinations (excluding dimer interactions inside the trimers),
%
%
and so on for the 4-, 5-, and 6-body interactions.
We have performed such a many-body decomposition for
the prism and cyclic conformers, since the prism conformer is favored
by the wave function approaches and the cyclic conformer is favored by many
of the DFT xc functionals.
The decomposition, the results of which are reported in Table \ref{tb3}, has been performed with MP2 (with an aug-cc-pV5Z basis set)
and with the X3LYP, PBE0, and BLYP xc functionals.
To enable an exact comparison between MP2 and the various XC functionals, absent
of any contributions arising from the slightly different structures obtained with
the different approaches, we
have used the MP2 geometries for all decompositions.\\

\begin{table*}
\caption {\label{tb3} Many-body contributions to the total dissociation energies
of the cyclic and prism isomers as obtained from MP2, X3LYP, PBE0, BLYP, and
BLYP+vdW.
For the MP2 many-body decomposition an aug-cc-pV5Z basis set is employed and so the
total MP2 dissociation energies
differ slightly from the MP2/CBS
%
values given in Table \ref{tb1}. Likewise, to avoid complications
from the slightly different optimized structures obtained from MP2 and the DFT xc functionals,
the DFT many-body decompositions are performed on the optimized
MP2 structures (with an aug-cc-pV5Z basis set for the DFT energies).
Values in the parenthesis are the difference between each functional and the MP2 results.
Negative values indicate a gain in energy, i.e., a net attraction when all
the n-body interactions of a given class are summed up, and positive values a net repulsion.
All values are in meV/H$_2$O.}
\begin{ruledtabular}
\begin{tabular}{cccccc}
           &    \multicolumn{5}{c} {\textbf{Cyclic}}   \\
\hline
           &      MP2   &   X3LYP  &  PBE0   & BLYP  & BLYP+vdW  \\
\hline
1-body      &     +16.6 &  +12.9 (-3.7) &  +16.5 (-0.1) & +2.4 (-14.2)  & +2.0 (-14.6)\\
2-body      &    -244.2 & -231.2 (+13.0)& -240.8 (+3.4) & -175.8 (+68.4)& -227.8 (+16.4)\\
3-body      &     -83.6 &  -92.1 (-8.5) &  -92.8 (-9.2) & -97.7 (-14.1) & -97.7 (-14.1)\\
4-body      &     -16.0 &  -13.9 (+2.1) &   -8.1 (+7.9) & -14.8 (+1.2)  & -14.8 (+1.2)\\
5-body      &      +0.5 &   -1.7 (-2.2) &   -6.4 (-6.9) & -1.9 (-2.4)   & -1.9 (-2.4)\\
6-body      &      -0.9 &   +0.0 (+0.9) &   +1.2 (+2.1) & +0.0 (+0.9)   & +0.0 (+0.9)\\
Total       &     -327.6&  -326.0 (+1.6)&  -330.4 (-2.8)& -287.8 (+39.8)& -340.2 (-12.6)\\
\hline
           &      \multicolumn{5}{c} {\textbf{Prism}} \\
\hline
           &      MP2   &   X3LYP &   PBE0   & BLYP & BLYP+vdW\\
\hline
1-body      &     +16.7 &  +14.4 (-2.3) &  +17.3 (+0.6) & +3.4 (-13.3)  & +3.2 (-13.5)\\
2-body      &    -283.4 & -263.6 (+19.8)& -274.4 (+9.0) & -191.8 (+91.6)& -278.0 (+5.4)\\
3-body      &     -63.8 &  -61.3 (+2.5) &  -59.3 (+4.5) & -79.3 (-15.5) & -79.3 (-15.5)\\
4-body      &      -5.2 &   -7.6 (-2.4) &   -5.2 (0.0)  & -2.8 (+2.4)   & -2.8 (+2.4)\\
5-body      &      -2.6 &   +1.4 (+4.0) &   -3.7 (-1.1) & +0.1 (+2.7)   & +0.1 (+2.7)\\
6-body      &      +2.2 &   -0.1 (-2.3) &   +2.5 (+0.3) & +0.1 (-2.1)   & +0.1 (-2.1)\\
Total       &    -336.1 & -316.8 (+19.3)& -322.8 (+13.3)& -270.3 (+65.8)& -356.7 (-20.6)\\
\end{tabular}
\end{ruledtabular}
\end{table*}

Let us first consider the MP2 reference data.
For each cluster a small positive 1-body
energy of $\sim$17 meV/H$_2$O is observed.
The 2-body interaction is attractive (negative) and at --244 meV
and --283 meV/H$_2$O for the cyclic and prism isomers, respectively,
comprises by far the largest contribution to the many-body expansion.
The 3-body interaction is also large and overall attractive:
--84 and --64 meV/H$_2$O for the cyclic and prism structures, respectively.
Indeed because of their magnitude the 2- and 3-body interactions almost decide
what the total dissociation energies are.
The 4-, 5-, and 6-body terms are all considerably smaller.
These results are consistent with those reported by Xantheas \textit{et al.}
\cite{xantheas_cooperativity} with a smaller basis set.\\

Turning our attention now to how the DFT xc functionals perform,
we first consider the two more accurate xc functionals for which the many-body
decomposition has been performed (PBE0 and X3LYP).
For the 1-, 4-, 5-, and 6-body contributions, we find
reasonably good agreement with MP2.
As we have said, these terms are small and the difference between MP2 and the two xc
functionals is typically $\ll$8 meV/H$_2$O.
For the (larger) 3-body terms we observe variable performance
with overbinding (8-9 meV) for the cyclic isomer and underbinding (3-5 meV) for the prism.
It is for the 2-body terms that we observe the largest deviations from MP2 with a consistent underbinding
for each functional and cluster.
%
%
Both PBE0 and X3LYP underestimate the
2-body contribution in the prism isomer by 9 and 20 meV/H$_2$O, respectively.
And for the cyclic isomer PBE0 and X3LYP underestimate the
2-body contribution by 4 and 13 meV/H$_2$O, respectively.
It is interesting that these errors are noticeably larger than the 1-2 meV/H$_2$O errors obtained
with these functionals for the equilibrium water dimer \cite{sms}.
Thus we observe from the many-body analysis that \emph{these xc functionals yield larger errors
when describing the non-equilibrium dimer configurations
present in the various water hexamers, compared to the equilibrium water dimer.}
Upon inspection of the errors associated with the individual dimer configurations within the hexamers
we find that there is a systematic underbinding for dimers at intermediate
separations (O-O distances $\sim$3.0 -- 5.0 \AA) typical of vdW bonded complexes
and also for certain orientations of water molecules held together with very non-linear H bonds.
There are not enough distinct dimer configurations within the hexamers to allow us to understand the precise
dependence of the 2-body error on orientation and H bond angle.
However, the distance dependence of the underbinding is more clear and is something that we now
address with a distance dependent vdW correction.
Before moving on we note that the BLYP errors from the many-body
analysis are consistently larger compared to PBE0 and X3LYP, consistent
with the generally inferior performance of this functional.
However, the main conclusion from the many-body analysis that the
2-body terms are underbound (and are more poorly described than the
equilibrium dimer) still holds. \\

%
%
\textbf{E. DFT+vdW DISSOCIATION ENERGY}\\

Nowadays it is well known that
%
most popular xc functionals generally
show unsatisfactory performance for
van der Waals forces, which inherently arise due
to non-local correlations \cite{Pulay-NoVdw,Becke-NoVdw,BenchmarkNonbonded}.
In order to test if the lack of van der Waals forces is indeed responsible
for the underestimation in the 2-body interactions, we use a simple $C_6R^{-6}$ correction for the DFT
total energies.
The $C_6R^{-6}$ correction method was early proposed for correcting HF calculations
\cite{Scoles1977},
and specifically applied to DFT by Wu and Yang \cite{wu_yang},
Grimme \cite{grimme_vdw} and
Jure\v{c}ka \emph{et al.} \cite{hobza_vdw}.
%
%
Certainly the $C_6R^{-6}$ scheme is a simple one for incorporating
dispersion interactions into DFT calculations
in contrast to other approaches
(e.g.
DFT xc functionals explicitly accounting for non-local correlation \cite{lundqvist_prl},
interaction of the instantaneous dipole moment of
the exchange hole \cite{becke_exchange_hole},
using maximally localized Wannier functions \cite{Silvestrelli_wannier}
or modified pseudopotentials \cite{anatole_prl}).
However, consistently accurate results have been obtained with the $C_6R^{-6}$
correction and it has a well established physical basis.
With this approach the pairwise vdW interaction ($E_{disp}$) is calculated by:
\begin{equation}
E_{disp} = - \sum_{j>i} f_{damp}(R_{ij},R_{ij}^0) C_{6ij} R_{ij}^{-6} \quad ,
\label{eqn_disp}
\end{equation}
where,
$C_{6ij}$ are the dispersion coefficients for an atom pair $ij$ (here taken from
the work of Wu and Yang \cite{wu_yang}),
$R_{ij}$ is the inter-atomic distance,
$R_{ij}^0$ is the sum of equilibrium vdW distances for the pair
(derived from atomic vdW radii \cite{vdw_radii_note}),
and $f_{damp}$ is a damping function.
The damping function is needed to avoid the divergence of the $R^{-6}$ term
at short distances and reduces
the effect of the correction on covalent bonds. We use a Fermi-type function $f_{damp}$,
%
\begin{equation}
f_{damp}(R_{ij},R_{ij}^0) = \left(1+\exp( -d (\frac{R_{ij}}{s_R R_{ij}^0} - 1 ))\right)^{-1} \quad ,
\label{eqn_damp}
\end{equation}
%
%
%
where, $d$ determines the steepness of the damping function
(the higher the value of $d$, the closer it is
to the step function),
and $s_R$ reflects the range of interaction covered by the
chosen DFT xc functional. The value of $d$ was set to 20 and
$s_R$ is 0.80 for BLYP, 1.00 for PBE and 1.03 for PBE0.
These values of d and $s_R$ were obtained by fitting on the intermolecular binding
energies of the S22 database \cite{hobza_vdw}
at the CBS limit for all DFT xc functionals
\cite{AT_vdW_new,vdw_Fit_note}.\\

\begin{table}
\caption {\label{tb4}
Absolute values of vdW interaction energies and vdW corrected total dissociation energies
for the four water hexamers for three different xc functionals. The DFT structures employed are fully relaxed geometries
calculated with the FHI-aims code (the CPMD code gives very similar numbers \cite{vdw_AIMS_note}).
For comparison the MP2/CBS results are also displayed.
The energies of the most stable isomers are indicated in bold
and the relative energies of the other structures with respect
to the prism are given in parenthesis.
MAE's in total dissociation energies are calculated from the MP2/CBS
values averaging over the
four hexamers. All numbers are in meV/H$_2$O.}
%
\begin{ruledtabular}
\begin{tabular}{cccccc}
& \multicolumn{5}{c}{\textbf{van der Waals interaction energy}}\\
\hline
Method & Prism & Cage & Book & Cyclic & \\
\hline
BLYP+vdW & 93.8 & 90.5 & 75.8  & 60.7 & \\
PBE+vdW  & 40.9 & 40.5 & 31.6  & 22.9 & \\
PBE0+vdW & 35.2 & 35.4 & 27.4  & 19.4 & \\
\hline
& \multicolumn{5}{c}{\textbf{Total dissociation energy}}\\
\hline
Method & Prism & Cage & Book & Cyclic & MAE\\
\hline
MP2 &\textbf{332.3}& 331.9(0.4) &  330.2(2.1) & 324.1(8.2)& ---\\
BLYP+vdW & \textbf{359.9} & 359.7(0.2)& 356.3(3.6)& 344.8(15.1)& 25.5\\
PBE+vdW  & 377.8 & \textbf{380.1}(-2.3) & 377.8(0.0)& 367.3(10.5)& 46.1\\
PBE0+vdW & 360.6 & \textbf{361.9}(-1.3) & 359.2(1.4)& 351.4(9.2)& 28.6\\
\end{tabular}
\end{ruledtabular}
\end{table}

The results for the PBE, PBE0 and BLYP functionals after applying the correction
on the four hexamers are shown in Table \ref{tb4}.
Also the total vdW interaction within each hexamer is reported.
One can see that the vdW correction is largest
for the prism and cage structures and noticeably less for book and cyclic structures;
favoring the prism or cage over the cyclic or the book structure.
The new energetic orderings of the hexamers are thus in contrast to all pure DFT functionals, which predict
the book or cyclic structures to have the lowest energy (Table \ref{tb1}), and
in better agreement with the wave function based methods.
The energy difference between the most stable
and the least stable hexamers is also in reasonably good agreement with MP2 and DMC results (around 10-15 meV).
Of the three functionals to which the correction has been applied, the BLYP+vdW method gives the
best agreement with MP2.
The MAE in the total dissociation energies for all four hexamers is reduced
from 15\% to 8\%.
And, moreover, the correct energetic ordering of the four isomers is recovered, i.e., BLYP+vdW
predicts the sequence prism $<$ cage $<$ book $<$ cyclic.
The results for BLYP+vdW are encouraging, however, it is important to
note
that there remains an 8\% error (a significant overbinding).
%
In addition,
the ``success'' of BLYP+vdW is achieved at the expense of a smaller $s_R$ parameter
which shifts the vdW minima to quite short distances (see below). Also, the three-body contribution
of BLYP, unaffected by the pairwise vdW correction, shows substantial error.
Thus, further investigation is required to rule out fortuitous error cancellation for BLYP+vdW.
Nonetheless these findings for at least three different functionals support the suggestion that the origin of the
incorrect prediction of the energetic ordering of the water hexamers
lies in the absence of vdW dispersion forces in the functionals
considered.\\

\textbf{IV. DISCUSSION AND CONCLUSIONS}\\

Having presented a lot of data obtained with various approaches, let
us now recap the main results and discuss them in a somewhat broader context.
To begin, there is the reference data itself, which has been acquired with
MP2 and diffusion quantum Monte Carlo (DMC).
From this we conclude that the prism is the lowest total energy isomer for six water molecules
in the absence of contributions from zero point vibrations.
This conclusion
agrees with the general consensus that has emerged, being
consistent with the very recent triple-$\zeta$ CCSD(T) results
\cite{hexamer_ccsdt, Truhlar_hexamer}.
There remain, of course, minor differences in the relative
energetic ordering of some structures on the order of 5 meV/H$_2$O
[notably CCSD(T) predicts particularly unstable book and cyclic structures compared to MP2,
with DMC being in between].
Resolving such small remaining differences will provide interesting work for the future.
In this regard CCSD(T) calculations
at the CBS limit would be welcome.
We stress that the ordering arrived
at here, prism $<$ cage $<$ book $<$ cyclic, is the ordering obtained in the absence
of corrections for zero point contributions.
It is known that zero point energies will alter the relative energy spacings with
indications that the cage becomes the most stable isomer
\cite{gregory_clary_96, gregory_clary_97, kim_kim_hexamer}.\\

%
%
It is interesting to see that
DMC and MP2 dissociation energies of the different isomers are so similar to each other,
within $4~$meV/H$_2$O.
This may indicate that correlation effects beyond MP2 have little effect on the hydrogen
bond energetics in these water clusters or it may indicate a favorable cancelation of errors in
the MP2 and/or DMC calculations.
%
%
Nonetheless, it demonstrates that DMC can achieve high accuracy
in describing the energetics of hydrogen bonds between water molecules,
already at the simplest DMC level, i.e. pseudopotential fixed-node DMC
with a single-determinant Slater-Jastrow trial wave function, as has been
found for a number of other hydrogen bonded systems (including DNA base
pairs)~\cite{korth08,fuchs05}. For the water hexamers studied here, the
fixed-node and pseudopotential approximation in DMC incur no significant
errors on the calculated hydrogen bond energies.
We stress, however, that in general such errors depend on the system considered
and still need to be carefully assessed by comparing to standard quantum chemistry
approaches such as e.g. CCSD(T)/CBS and monitoring the quality of the trial
wave function and, when used, also the pseudopotentials.\\

%
%
The main part of this paper was concerned with
using the reference data from the wave function based methods to evaluate the performance of
several DFT xc functionals.
A sub-set of the xc functionals previously tested for
small water clusters \cite{sms} was considered.
It was found that whilst certain functionals did a reasonable job at predicting the absolute
dissociation energies of the various isomers (coming within 10-20 meV/H$_2$O),
\emph{none} of the functionals tested predict the correct energetic ordering of the four isomers,
nor does any predict the correct lowest energy isomer.
All xc functionals either predict the book or cyclic isomers
to have the largest dissociation energies.
There have been indications before that certain DFT xc functionals may not predict
the correct lowest energy structure for the water hexamer.
%
BLYP, for example, was long
ago shown to favor the cyclic isomer \cite{fitzgerald_hexamer}.
Likewise X3LYP, B3LYP, and PBE1W have been shown to favor
the cyclic structure \cite{X3LYP_water, Truhlar_hexamer}.
Here, we have shown that several other popular xc functionals fail
to predict the correct lowest energy structure too.
Furthermore, by attributing the failure to an improper
treatment of vdW forces it seems likely that many
other semi-local and hybrid xc functionals which do not account for vdW
in some way will also fail in this regard.
We have shown that by augmenting the BLYP functional
with an empirical pairwise $C_{6}R^{-6}$ correction the
correct energetic ordering of the four hexamers is recovered.
Equivalent empirical corrections to other functionals (PBE, PBE0) also
improves the ordering somewhat, favoring the prism and cage isomers
over the book and cyclic ones.
Of course there are other means of incorporating vdW dispersion forces implicitly
into DFT xc functionals such as the approaches pioneered by Lundqvist and Langreth
and Becke and others
\cite{lundqvist_prl, becke_exchange_hole, Silvestrelli_wannier, anatole_prl}.
It will be interesting to see if these functionals
can predict the correct lowest energy structure for the water hexamer and, at
the same time, yield accurate total dissociation energies.
Indeed on the general point of benchmarking and accessing the performance
of existing and new xc functionals for the treatment of H bonded systems,
it seems that the water hexamer would be an appropriate test case to add
to existing H bond test sets since it presents a stern challenge for
any xc functional.
We reiterate that we are not suggesting that all xc functionals which do not account for
vdW forces in one way or another are likely to fail to predict the correct
energy ordering for the water hexamer.
Indeed Truhlar and co-workers have very
recently reported that a few empirical hybrid meta-GGA functionals achieve the
correct energetic ordering \cite{Truhlar_hexamer, M05-2X_note}.
This looks like an exciting development but what the
precise reason for the success of the functionals tested is remains unclear to us at present.
\\


\begin{figure}
   \begin{center}
       \includegraphics[width=8cm]{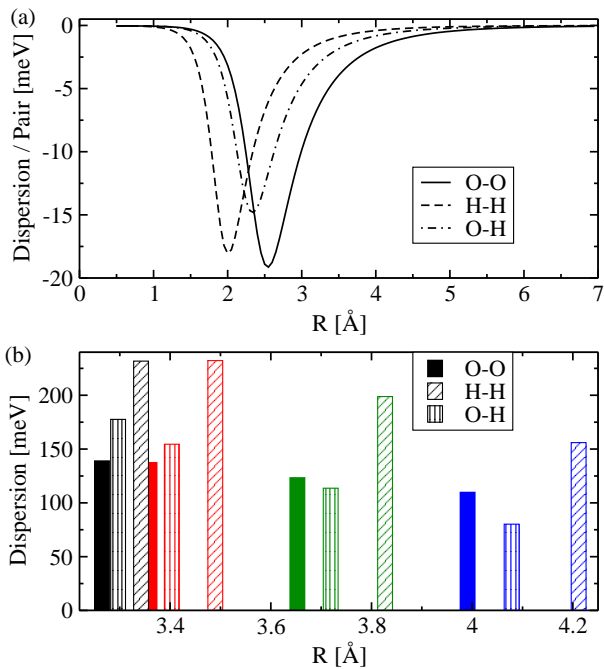}
   \end{center}
   \caption{\label{fig4} (a) Variation in the dispersion contribution
   with distance from different atom pairs with parameters for
   BLYP.
   (b) Inter-molecular dispersion interaction for the four isomers as a
   function of the average inter-atomic distances of different atom pairs
   (on BLYP+vdW optimized structures).
   Here black, red, green, and blue refer to prism, cage, book,
   and cyclic isomers, respectively.}
\end{figure}

Having identified a lack of vdW dispersion forces as being at
the heart of the incorrect energy ordering of the various water hexamers, we now consider
why the $C_{6}R^{-6}$ correction scheme applied here works to alter
the relative energies of the four isomers.
Since the BLYP+vdW scheme performs
reasonably well
for this system and recovers the correct energetic ordering for
the four hexamers we focus on analyzing the details of this correction.
First we consider the functional form of the
dispersion corrections applied in these systems.
These are displayed in Fig. \ref{fig4}(a) for the three individual types of
atom-atom interaction: O--O, O--H, and H--H.
%
%
Dispersion forces are generally considered to be long range
and indeed the tails of all three vdW curves extend to beyond
4 \AA.
However, the minima of the vdW curves are located at considerably
shorter distances: $\sim$2.80, $\sim$2.20, and $\sim$2.55 \AA\ for
the O--O, H--H, and O--H curves, respectively.
It is the location of these vdW minima relative to the structures
of the various isomers that leads to the revised energetic reordering of the
four isomers.
In simplest terms the mean inter-molecular distances of the
four clusters decreases upon going from cyclic to
book to cage to prism and so the magnitude of the dispersion correction
decreases in the order prism to cage to book to cyclic, which ultimately
leads to the correct stability sequence prism to cage to book to cyclic.
Considering this in more detail we show in Fig. \ref{fig4}(b) the contributions to the total
inter-molecular dispersion interaction in each cluster for each type of atomic
pair interaction (O--O, H--H, and O--H), plotted as a function
of distance \cite{fig4_note}.
It can be seen from the histogram that the average inter-molecular O--O, O--H, and H--H
distances steadily increase along the sequence prism-cage-book-cyclic and that likewise the
dispersion contribution decreases.
Further, we note that by simply summing up the contributions from each type of interaction
in the hexamers
we find that the majority of the vdW correction comes from H-H interactions ($\sim$44-48\%),
followed by the O-H ($\sim$22-32\%)
and then the O-O ($\sim$25-30\%) interactions.
The H--H interaction dominates simply because there are more them.
For brevity we do not show the results of similar analysis performed for
the PBE and PBE0 vdW corrections.
However, the general conclusion that the vdW dispersion contribution favors
the more compact prism and cage isomers over the less compact
book and cyclic isomers because the former are closer to the minima
of the vdW curves than the latter also
holds for the PBE and PBE0 vdW corrections.\\

Finally, this paper has focused on water clusters. However, it does not
seem unreasonable to suggest that the results presented here
will be of some relevance to DFT simulations of liquid water.
Certainly if an xc functional encounters difficulties in predicting the
correct energetic ordering of the low energy isomers of the water hexamer
it is likely that similar errors will exist in describing the
many more competing configurations of water clusters present transiently
or otherwise in the liquid.
Given that the hybrid xc functionals PBE0 and X3LYP also fail for the hexamer,
despite otherwise predicting H bond strengths and structures for
smaller water clusters in excellent agreement with MP2 it
seems likely that these functionals may  not
offer the promise anticipated for liquid water \cite{sms}.
Indeed a very recent PBE0 simulation for liquid water which ran for
a reasonably respectable 10 ps, found that the PBE
and PBE0 RDFs were essentially indistinguishable \cite{vandevondele_08}.
%
Based on the forgoing results and discussion the lack of a significant improvement
in describing the liquid is not entirely unexpected.
We suggest instead that density-functional
methodologies which account for vdW dispersion forces are likely
to offer more promise in the quest to improve the description of
liquid water.
Again very recent MD simulations of liquid water are consistent with this
suggestion. Lin \emph{et al.} have reported BLYP simulations for liquid water corrected
with a similar C$_6$R$^{-6}$ correction scheme to the one employed here
(but with a different damping function)
as well as a separate account for vdW through the use of modified pseudopotentials
\cite{Lin_vdw_water_new}.
These simulations indicate that (at the experimental density and temperatures tested)
accounting for vdW forces lowers the peak maximum in the O-O RDF, and in
so doing brings the experimental and theoretical RDFs into better agreement. \\

\textbf{Acknowledgments}\\

This work is supported by the European Commission through the Early
Stage Researcher Training Network MONET, MEST-CT-2005-020908 (\emph{www.sljus.lu.se/monet}).
A.T. thanks the Alexander von Humboldt (AvH) Foundation for funding.
A.M's work is supported by a EURYI award (\emph{www.esf.org/euryi})
and by the EPSRC.\\


\begin{thebibliography}{115}
\bibitem[{not({\natexlab{a}})}]{note_g03_nwchem}
\bibinfo{note}{We have used Gaussian03 \cite{g03} and NWChem \cite{nwchem}
  interchangeably, since total energies of the water clusters obtained from the
  two codes differ by no more than 0.4 meV per water molecule.}
\bibitem[{ext()}]{extrap_scheme_comment}
\bibinfo{note}{As we have said before \cite{sms}, we opted for the scheme
  provided by eqns. (\ref{eqn_HF_extrap}-\ref{eqn_Correlation_extrap}) because
  we found that with input from triple-, quadruple-, and pentuple-$\zeta$ basis
  sets this method was best able to predict the total energy of a water monomer
  and dimer explicitly calculated with an aug-cc-pV6Z basis set (443 basis
  functions/H$_2$O).}
\bibitem[{mp2()}]{mp2_error}
\bibinfo{note}{It is somewhat difficult to determine the precise ``error bar''
  associated with our MP2/CBS estimates. However, as we noted before through
  comparisons of MP2/CBS and CCSD(T)/CBS, the error [due to MP2 (valence only)
  treatment of correlation, aug-cc-pVTZ structures, extrapolation to reach the
  CBS, etc.] is not likely to be more than 5 meV/H bond.}
\bibitem[{not({\natexlab{c}})}]{note_CCSD(T)}
\bibinfo{note}{We believe the differences between CCSD(T) and MP2 (10-15
  meV/H$_2$O) come mostly from the use of an aug-cc-pVTZ basis set for the
  CCSD(T) energy calculations. Indeed with this basis set the difference
  between MP2 and CCSD(T) is reduced to only 4 meV/H$_2$O. Specifically, our
  MP2/aug-cc-pVTZ dissociation energies are 346.5, 345.9, 343.0, and 335.1
  meV/H$_2$O, respectively, for prism, cage, book, and cyclic isomers. In
  addition, the different basis sets used to obtain the structures is likely to
  account for some of the remaining difference. Here we use MP2 structures
  optimized with an aug-cc-pVTZ basis set where as in ref. \cite{hexamer_ccsdt}
  MP2 structures optimized with a polarized double-$\zeta$ basis set of Dunning
  and Hay \cite{dunning_hay}, DH(d,p)] are used.}
\bibitem[{not({\natexlab{d}})}]{note_on_small_clusters}
\bibinfo{note}{Dissociation energies calculated (using BLYP) respectively with
  G03/aug-cc-pV5Z, CPMD, and FHI-aims codes are the following: dimer (180.7,
  179.3, 180.6), trimer (191.7, 190.8, 191.9), tetramer (264.9, 263.9, 265.1),
  and pentamer (281.2, 280.7, 281.3). Likewise in comparison to G03/aug-cc-pVTZ
  structures CPMD and FHI-aims respectively concede maximum error of 0.031,
  0.009 \AA\ for bond lengths and 0.57$^\circ$, 0.03$^\circ$ for bond angles
  (parameters considered are shown in Table \ref{tb2}).}
\bibitem[{vdw({\natexlab{a}})}]{vdw_radii_note}
\bibinfo{note}{We have used Bondi's vdW radii for different atoms
  \cite{bondi_radii}. More precisely for the atom pairs we used the following
  radii and vdW coefficients (in a.u.): $R_{H-H}^0=4.54$, $R_{O-O}^0=5.74$,
  $R_{O-H}^0=5.28$, $C_{6H-H}=2.78$, $C_{6O-O}=12.14$, and $C_{6O-H}=5.67$. For
  more details on the averaging of vdW radii and dispersion coefficients for
  pairs of atoms see Ref. \cite{hobza_vdw}.}
\bibitem[{vdw({\natexlab{b}})}]{vdw_Fit_note}
\bibinfo{note}{Also we made fits on various water dimers to reproduce the MP2
  2-body interaction by optimizing the correction parameters using BLYP. A
  large set of parameters was considered by a careful scan over the range
  $0.8\le s_R\le 1.5$ and $20\le d\le 40$. All possible combinations of water
  dimers from all four hexamers (total 48 dimers) were included in the fitting
  and interestingly the set of parameters which yielded the smallest MAE for
  the 48 dimers are $s_R=0.82$ and $d=20$, which are nearly the same as those
  found from the S22 databse.}
\bibitem[{vdw({\natexlab{c}})}]{vdw_AIMS_note}
\bibinfo{note}{We have also performed full geometry optimizations with vdW
  corrected BLYP and PBE using the CPMD code and the total dissociation
  energies are within 1 meV from the numbers presented in Table \ref{tb3}. We
  also note that with FHI-aims code PBE0+vdW calculations were performed on
  PBE+vdW optimized geometries.}
\bibitem[{M05()}]{M05-2X_note}
\bibinfo{note}{Our calculations with the M05-2X functional with a larger basis
  set (aug-cc-pVQZ) also find the prism to be the lowest energy structure.
  Specifically, the dissociation energies are 353.5, 347.9, 339.9, and 333.1
  meV/H$_2$O for prism, cage, book, and cyclic, respectively.}
\bibitem[{fig()}]{fig4_note}
\bibinfo{note}{The total intra-molecular O-H and H-H dispersion contributions
  are 0.6 and $\sim13-17$ meV, respectively. These are small compared to
  intermolecular contributions and remain almost constant for all the hexamers
  and for brevity are not shown in Fig. \ref{fig4}.}
\end{thebibliography}

\end{document}